\documentstyle[prl,aps,epsf,floats]{revtex}
\begin{document}
\draft
\twocolumn[\hsize\textwidth\columnwidth\hsize\csname @twocolumnfalse\endcsname
\title {Continuous-Time Quantum Monte Carlo Algorithm for 
        the Lattice Polaron} 

\author{P.\,E.\,Kornilovitch}
\address{
Blackett Laboratory, Imperial College, Prince Consort Road, \\ 
London SW7 2BZ, United Kingdom}

\date{17 August 1998}
\maketitle
\begin{abstract}

An efficient continuous-time path-integral Quantum Monte 
Carlo algorithm for the
lattice polaron is presented. It is based on Feynman's
integration of phonons and subsequent simulation of the 
resulting single-particle self-interacting system. The 
method is free from the finite-size and finite-time-step 
errors and works in any dimensionality and for any range 
of electron-phonon interaction. The ground-state energy
and effective mass of the polaron are calculated  
for several models. The polaron spectrum can be measured
directly by Monte Carlo, which is of general interest. 

\end{abstract}
\vskip2pc]
\narrowtext

The past few years have witnessed a rapid development of 
continuous-time
Quantum Monte Carlo (QMC) algorithms for quantum-mechanical
lattice models. The driving force behind this
is the desire to eliminate the systematic errors introduced
by the Trotter decomposition in the standard discrete-time
QMC methods \cite{DeRaedt_one}. The main idea then is to regard the
imaginary-time evolution of a particle or spin configuration
as a continuous-time Poisson process with ``events'' being
either a particle jump or a spin flip. 
In this way continuous-time QMC 
algorithms were developed for a particle in an external potential 
\cite{Prokofiev}, Heisenberg model \cite{Beard,Kashurnikov},
$t-J$ model \cite{Ammon}, bosonic Hubbard model \cite{Kashurnikov_two},
and Fr\"ohlich polaron \cite{Prokofiev_two}.

In this Letter I present a continuous-time path-integral QMC 
algorithm for the lattice polaron, i.e., an electron strongly 
interacting with phonons on a lattice. The method combines 
analytical integration of the phonon degrees of freedom 
with the advantages of the continuous-time formulation of the
Monte Carlo process. The method is universal. It works
for infinite lattices in any dimensionality and for any 
radius of electron-phonon interaction. It is also free from 
the systematic finite-time-step errors which were an undesirable 
feature of the original (discrete-time) QMC algorithm based 
on the integration of phonons \cite{DeRaedt_two}. In fact, the
new method allows for {\em exact} (in the QMC sense) calculation
of the ground-state energy, effective mass, and even 
{\em spectrum} of the polaron. The numerical accuracy
of the method is $0.1 - 0.3 \%$, which is not as good as that
of exact diagonalization \cite{Kabanov_one,Fehske} and density-matrix
renormalization group \cite{White} schemes, but is good enough
for practical purposes. 

I begin with the important question of which quantities
can be calculated with a path-integral QMC method.
The traditional scheme samples paths which are periodic
in imaginary time, see, e.g., Refs.~\cite{DeRaedt_one,DeRaedt_two}.
This allows for the calculation of thermodynamic properties such
as internal energy or specific heat, as well as some static
correlation functions, but not dynamic properties of the system.
In Ref.\cite{Kornilovitch} it was shown that the polaron effective 
mass, an important dynamic characteristic, can be measured
on the ensemble of paths with {\em open} boundary conditions
in imaginary time (BCIT). Here I extend the argument
to the whole polaron spectrum. Consider two quantities.
The first one, $Z_{\bf P} = \sum_i \langle i | e^{-\beta H}
| i \rangle \delta_{{\bf P}_i, {\bf P}}$, is by definition
the partition function of many-body states with fixed 
total (electron plus phonon) quasimomentum ${\bf P}$
[$\beta=(k_B T)^{-1}$ and $H$ is the Hamiltonian]. 
The second quantity is the partition function with twisted BCIT,
$Z_{\triangle {\bf r}} = {\rm Tr}_{\triangle {\bf r}}{e^{-\beta H}}$,
in which the many-body configurations (electron position ${\bf r}$  
and ionic displacements $\vec \xi$) at $\tau=\beta$ are obtained
from the configurations at $\tau = 0$ by shifting along
the lattice by the vector $\triangle {\bf r}$. There exists a
fundamental Fourier-type relation between the two 
\cite{Kornilovitch}: 
\begin{equation}
Z_{\bf P} = \sum_{\triangle {\bf r}} 
e^{i {\bf P} \triangle {\bf r} }  Z_{\triangle {\bf r}} =
\int_{op} {\cal D} {\bf r} {\cal D} \vec \xi \,  
e^{i {\bf P} \triangle {\bf r} } 
w[ {\bf r}(\tau) \vec \xi(\tau) ] .
\label{one}
\end{equation}
Here $w[{\bf r}(\tau) \vec \xi(\tau) ]$ is the (positive-definite)
weight of a many-body path satisfying the twisted BCIT, and
$\int_{op} {\cal D} {\bf r} {\cal D} \vec \xi $ 
means the integration over such 
paths with all possible shifts $\triangle {\bf r}$.
In the low-temperature limit, $\beta \rightarrow \infty$,
$Z_{\bf P}$ is dominated by the lowest eigenstate $E_{\bf P}$:
$Z_{\bf P}$$=$$\exp (-\beta E_{\bf P})$ and 
$E_{\bf P}$$=$$-\frac{1}{Z_{\bf P}} 
\frac{\partial Z_{\bf P}}{\partial \beta}$.
Equation (\ref{one}) then gives
\begin{equation}
E_{\bf P} = 
\frac{\int_{op} {\cal D} {\bf r} {\cal D} \vec \xi \, 
e^{i {\bf P} \triangle {\bf r} }
\left[ - \frac{1}{w} \frac{\partial w}{\partial \beta} \right] w}
{\int_{op} {\cal D} {\bf r} {\cal D} \vec \xi \, 
e^{i {\bf P} \triangle {\bf r} } w}  
\equiv 
\frac{\langle -e^{i {\bf P} \triangle {\bf r} } 
\frac{1}{w} \frac{\partial w}{\partial \beta} \rangle_0}
{\langle e^{i {\bf P} \triangle {\bf r} } \rangle_0} ,
\label{two}
\end{equation}
where $\langle A \rangle_0 \equiv 
(\int_{op}{\cal D}{\bf r} {\cal D} \vec \xi \, w)^{-1}
\int_{op} {\cal D}{\bf r} {\cal D} \vec \xi \, A w $ 
stands for the average in the case ${\bf P}=0$. 
Equation (\ref{two}) shows that the
ground-state dispersion of the polaron can be inferred
{\em directly} from imaginary-time simulations. 
However, the formalism is not free from the
sign-problem, as apparent from Eq. (\ref{two}).
Energies for non-zero values of ${\bf P}$ can be obtained only
at intermediate and large electron-phonon couplings
when the average  
$\langle e^{ i {\bf P} \triangle {\bf r} } \rangle_0
= \langle \cos{ {\bf P} \triangle {\bf r} } \rangle_0$ is not
small. The ground state corresponds 
to ${\bf P}=0$ and its energy can be calculated 
straightforwardly using Eq. (\ref{two}). 

Analogous formulas can be obtained for derivatives of $E_{\bf P}$
with respect to momentum, e.g., for the inverse effective mass.
In the low-temperature limit one can write
$E_{\bf P} = - \frac{1}{\beta} \ln Z_{\bf P}$ since there is no
difference between $1/\beta$ and $\partial/ \partial \beta$. 
From this and Eq.~(\ref{one}), it follows that
\begin{equation}
\frac{1}{m^{\ast}_{\alpha}} \! \equiv \! 
\left. \frac{1}{\hbar^2} \frac{\partial^2 E_{\bf P}}
{\partial P^2_{\alpha}}\right\vert_{{\bf P}=0} \!\!\!\!\!\!\!\! = \!
\frac{\int_{op} {\cal D} {\bf r} {\cal D} \vec \xi  
\left[ \frac{(\triangle r_{\alpha})^2}{\hbar^2 \beta} \right] \! w }
{\int_{op} {\cal D} {\bf r} {\cal D} \vec \xi \, w}  = 
\frac{\langle (\triangle r_{\alpha})^2 \rangle_0}{\hbar^2\beta} . 
\label{three}
\end{equation}
One can see that the inverse effective mass is the 
``diffusion coefficient'' of the imaginary-time propagation. 
(For similar treatment of the continuous polaron and of an atom 
of $^3$He in liquid $^4$He, see \cite{Alexandrou}
and \cite{Ceperley}, respectively.)

Thus, to compute dynamical properties, the QMC process should 
sample paths with open BCIT. 
Consider now a particular polaron system on a
hypercubic lattice with nearest-neighbor hopping.
The model is defined by the Hamiltonian
\begin{equation}
H = -t \sum_{\bf nn'} c^{\dagger}_{\bf n'} c_{\bf n} 
- \sum_{\bf mn} f_{\bf m}({\bf n}) c^{\dagger}_{\bf n} c_{\bf n} 
\xi_{\bf m} + \hbar \omega \sum_{\bf m} 
b^{\dagger}_{\bf m} b_{\bf m} . 
\label{four}
\end{equation}
The phonon subsystem (operators $b^{\dagger}_{\bf m}, b_{\bf m}$)
is a set of uncoupled harmonic oscillators, one per 
site, with internal coordinates $\xi_{\bf m}$, frequency
$\omega$, and reduced mass $M$. The lattice is 
assumed to be infinite in all dimensions. The electron-phonon 
interaction is taken to be of the ``density-displacement'' form. 
No restriction is imposed on the form of  
{\em force} $f_{\bf m} ({\bf n})$ with which the electron at
site ${\bf n}$ acts on ${\bf m}$'th oscillator. 

Since Feynman's classic work on polarons \cite{Feynman},
it has been known that phonon degrees of freedom 
(variables $\vec \xi(\tau)$) can be integrated out in the 
path-integral representation of the partition function, 
Eq.~(\ref{one}), leading to a retarded self-interaction 
of the electron. For {\em periodic} BCIT, 
$\xi_{\bf m}(\beta) = \xi_{\bf m}(0)$,
the phonon-induced part of the polaron action is \cite{Feynman}
%
%\begin{equation}
%A_{per}[ {\bf r}(\tau) ] = \frac{\hbar}{4M\omega}
%\int^{\beta}_0 \!\!\! \int^{\beta}_0 \!\!\!
%d\tau_1 d\tau_2 \frac{\cosh \hbar\omega (\frac{\beta}{2} - 
%|\tau_1-\tau_2|)} {\sinh \frac{\hbar \omega \beta}{2}} 
%\sum_{\bf m} f_{\bf m}({\bf r}(\tau_1)) f_{\bf m}({\bf r}(\tau_2)) , 
%\label{five}
%\end{equation}
%
\begin{displaymath}
A_{per}[ {\bf r}(\tau) ] = \frac{\hbar}{4M\omega}
\int^{\beta}_0 \!\!\! \int^{\beta}_0 \!\!\!
d\tau_1 d\tau_2 \frac{\cosh \hbar\omega (\frac{\beta}{2} - 
|\tau_1-\tau_2|)} {\sinh \frac{\hbar \omega \beta}{2}} \times
\end{displaymath}
\vspace{-0.5cm}
\begin{equation}  
\sum_{\bf m} f_{\bf m}({\bf r}(\tau_1)) f_{\bf m}({\bf r}(\tau_2)) , 
\label{five}
\end{equation}
where ${\bf r}(\tau)$ is the electron path. However,
in the case of open BCIT the integration must be performed
under the constraint 
$\xi_{\bf m + \triangle r}(\beta) = \xi_{\bf m}(0)$, where
$\triangle {\bf r} = {\bf r}(\beta) - {\bf r}(0)$ is
the shift of the electron path. 
For nonzero $\triangle {\bf r}$ this yields an {\em extra term} 
in the action
\begin{equation}
\triangle A [ {\bf r}(\tau) ] =  \frac{\hbar}{2M\omega}
\sum_{\bf m} B_{\bf m} ( C_{\bf m + \triangle r} - C_{\bf m}) ,
\label{six}
\end{equation}
where
\begin{equation}
B_{\bf m} \! = \! \int^{\beta}_0 \!\! 
d\tau e^{-\hbar\omega\tau} \!f_{\bf m}({\bf r}(\tau)), \; 
C_{\bf m} \! = \! \int^{\beta}_0 \!\!  
d\tau e^{-\hbar\omega(\beta-\tau)} \! f_{\bf m}({\bf r}(\tau)). 
\label{seven}
\end{equation}
This formula is valid only in the limit $e^{\beta \hbar \omega} \gg 1$, 
which is easy to realize in practice.
Thus, the integration over phonons reduces the problem to 
a single-particle system with an extra factor 
$\exp(A) = \exp( A_{per} + \triangle A )$ in the weight
of each path ${\bf r}(\tau)$. 

\begin{figure}[t]
\begin{center}
\leavevmode
\hbox{
\epsfxsize=8.4cm
\epsffile{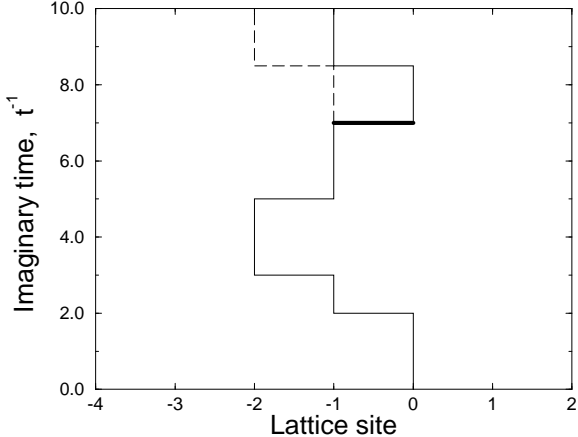}
}
\end{center}
\vspace{-0.5cm}
\caption{ 
A typical polaron path for $\beta = 10\,t^{-1}$ (solid line).
When the kink at $\tau_0=7.0\,t^{-1}$ (bold line) is removed, 
the whole path at $\tau_0 < \tau < \beta$ is shifted to the
left by one lattice site (dashed line). 
$\triangle r$ changes from -1 to -2.
}
\label{fig1}
\end{figure}

I now describe how this factor should be incorporated
into the general scheme of continuous-time QMC \cite{Prokofiev_three}. 
In $d$ dimensions the hopping term in the Hamiltonian
(\ref{four}) introduces $2d$ {\em independent} Poisson
processes with events, or ``kinks'' (after Ref. 
\cite{Prokofiev}), being jumps of an electron path to
one of $2d$ nearest neighbors. The probability to have
$N_k$ kinks of a given sort on the time interval $[0,\beta]$
is given by the Poisson distribution
$P_{N_k} = (N_k!)^{-1} (t\beta)^{N_k} e^{-t\beta}$.
The Monte Carlo process consists of (i) proposing a change 
of the path by either removing existing kinks or adding
new kinks and (ii) either accepting or rejecting the proposal.
(Another possible subprocess is the shift of a kink in time 
but this can always be achieved by adding and removing.) 
Because of the open BCIT, it is sufficient to change the number 
of kinks just by one. When a kink
is added (removed) at time $\tau_0$ the {\em whole} path
at $\tau > \tau_0$ is shifted in the corresponding direction
(antidirection) by one lattice site (see Fig. \ref{fig1}). 
The balance equation for the adding-removing process is
%
%\begin{equation}
%q_a W(N_k) P_{a} (N_k \rightarrow N_k+1) =
%q_r W(N_k + 1) P_{r} (N_k+1 \rightarrow N_k) ,
%\label{eight}
%\end{equation}
%
\begin{displaymath}
q_a W(N_k) P_{a} (N_k \rightarrow N_k+1) = \makebox[2.cm]{}
\end{displaymath}
\vspace{-0.7cm}
\begin{equation}
\makebox[2.cm]{} 
q_r W(N_k + 1) P_{r} (N_k+1 \rightarrow N_k) ,
\label{eight}
\end{equation}
where $W(N_k)$ is the probability to have a given path with
$N_k$ kinks of a given sort, and $q_a$ and $q_r$ are the probabilities
of attempting to add or remove a kink, respectively.
In this paper, $q_a$$=$$q_r$$=$$1/2$ is used for $N_k$$\geq$$1$, and
$q_a$$=$$1$, $q_r$$=$$0$ for $N_k$$=$$0$ (if there are no 
kinks of a given sort, one can only add one). In the absence
of electron-phonon interaction the ratio 
$W(N_k+1)/W(N_k) = t\beta/(N_k+1)$ follows from the
Poisson distribution. In the general case, each $W$ is multiplied 
by its phonon-induced weight $e^A$. The acceptance rules now 
follow from Eq. (\ref{eight}):
\begin{equation}
P_{a} (N_k \rightarrow N_k+1) = {\rm min} \left[ 1,
\frac{t\beta}{N_k+1} e^{A_{N_k+1} - A_{N_k}} \right],
\label{nine}
\end{equation}
to add a new kink to the existing $N_k$ of a given sort, and
\begin{equation}
P_{r} (N_k+1 \rightarrow N_k) = {\rm min} \left[ 1,
\frac{N_k+1}{t\beta} e^{A_{N_k} - A_{N_k+1}} \right],
\label{ten}
\end{equation}
to remove one of the existing $N_k+1$ kinks of a given sort.
For the special case $N_k=0$, the preexponential factor in
Eq.~(\ref{nine}) must be $t\beta/2$ instead of $t\beta$,
and in Eq.~(\ref{ten}) it must be $2/(t\beta)$ instead
of $1/(t\beta)$.

The explicit form of the energy estimator that enters 
Eq.~(\ref{two}) follows from the $\triangle \tau \rightarrow 0$
limit of the corresponding finite-time expression \cite{DeRaedt_two}
\begin{equation}
-\frac{1}{w} \frac{\partial w}{\partial \beta} =
- \frac{N^{tot}_k}{\beta} - \frac{\partial A}{\partial \beta} ,
\label{eleven}
\end{equation}
where $N^{tot}_k$ is the total number of kinks (of all sorts)
on a path. Note that due to the open BCIT the thermodynamic
estimator (\ref{eleven}) measures the ground-state energy 
rather than full internal energy.

Summarising the procedure, the QMC process samples 
single-particle paths with open BCIT by inserting and deleting 
kinks of $2d$ types. The acceptance rules for these two fundamental 
processes are given by Eqs.~(\ref{nine}) and (\ref{ten}). The 
action $A$ is a functional of the electron
path ${\bf r}(\tau)$ and is given by the sum of Eqs.~(\ref{five})
and (\ref{six}). Measured quantities include the energy
estimator (\ref{eleven}) and the inverse mass estimator
$\frac{1}{\beta}(r_{\alpha}(\beta)$$-$$r_{\alpha}(0))^2$. The polaron
spectrum and effective mass are calculated with Eqs.~(\ref{two})
and (\ref{three}), respectively. Note that statistics for
{\em all} momenta (i.e., for as many ${\bf P}$-points as
one wants) are collected during a {\em single} QMC run. 
In practical simulations
successive measurements were taken every tenth single-kink
step to reduce statistical correlations. For each set of 
model parameters several series of 500\,000 or 1\,000\,000 
measurements were conducted for different values of $\beta$ 
to detect possible finite-temperature systematic errors, 
typically for $\beta \hbar \omega = 10, 15, 20,$ and 25. 
No such errors were detected.

\begin{figure}[t]
\begin{center}
\leavevmode
\hbox{
\epsfxsize=8.4cm
\epsffile{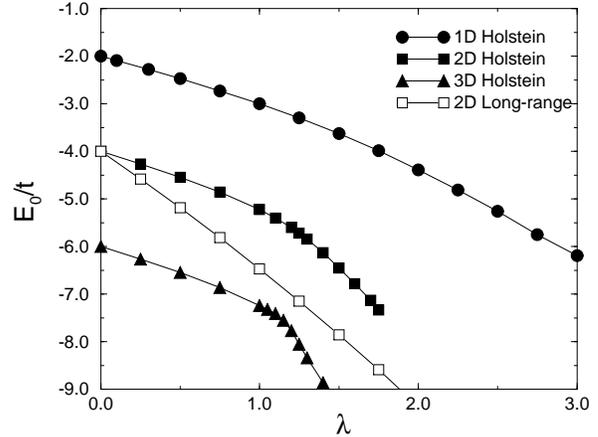}
}
\end{center}
\vspace{-0.5cm}
\caption{ 
Ground-state polaron energy for several models.
In all cases, $\bar\omega=1.0$.
Statistical errors are smaller than the symbols.
}
\label{fig11}
\end{figure}

\begin{figure}[t]
\begin{center}
\leavevmode
\hbox{
\epsfxsize=8.4cm
\epsffile{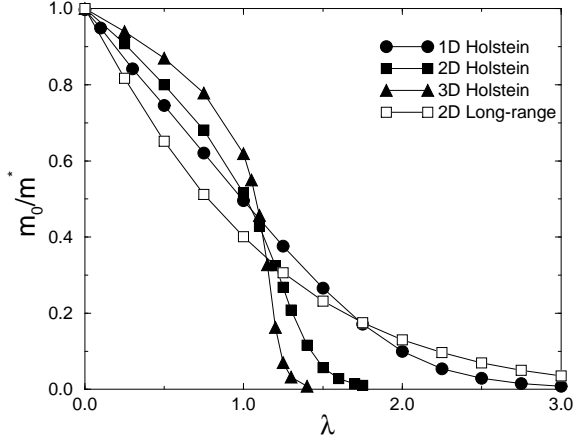}
}
\end{center}
\vspace{-0.5cm}
\caption{ 
Inverse effective polaron mass in units of
$1/m_0 = 2ta^2/\hbar^2$ for several models. In all cases,
$\bar\omega = 1.0$. Statistical errors are smaller than the
symbols.
}
\label{fig2}
\end{figure}

The method was tested on the Holstein model,
for which independent numerical results are available. The
Holstein model is a particular case of Eq.~(\ref{four}) with
$f_{\bf m}({\bf n})$$=$$\kappa \delta_{\bf mn}$. 
The model is parametrized by the di\-men\-sion\-less frequency
$\bar\omega$$=$$\hbar \omega/t$ and the dimensionless coupling constant 
$\lambda \equiv (\sum_{\bf m} f^2_{\bf m}(0))/(4dtM\omega^2) =
\kappa^2/(4dtM\omega^2)$. 
Excellent agreement with previously published or communicated 
results was found. For example, in $d=1$ [Figs.~\ref{fig11} and 
\ref{fig2} (circles)] the present method's estimate of the 
ground state energy for $\bar\omega=1.0, \lambda=0.5$ 
is $E_0= -2.471 \pm 0.001$ (in units of $t$), with the true value 
being in the interval (-2.46968, -2.471) \cite{Brown};
for $\bar\omega=1.0, \lambda=1.0$ it yields $E_0 = -2.999 \pm 0.001$
with the true value in  (-2.99883, -3.000) \cite{Brown}.  
For $\bar\omega=1.0$ and $\lambda=1.25$, 1.5, and 2.0, the
estimated energies are -3.298, -3.623, and -4.388, respectively, 
which are, to within statistical error ($\pm 0.002$), precisely 
the values obtained by the exact diagonalization (ED)
method \cite{Kabanov}. For $\bar\omega=2.0$ and
$\lambda = 1.5625$ and 2.25, QMC yields $E_0 = -4.013$ and
-5.070, respectively, in agreement with the 
strong-coupling perturbation theory \cite{Stephan}.
Polaron masses obtained by QMC are in excellent agreement 
with variational calculations \cite{Brown} and with 
density-matrix renormalization group (DMRG) results \cite{White}. 
In $d=2$, $\bar\omega=1.0$ [Figs.~\ref{fig11} and \ref{fig2} (filled 
squares)], QMC's energies agree with ED
\cite{Wellein_two} and QMC's masses agree with DMRG \cite{White}. 
All of these checks confirm that the present algorithm
is free from systematic errors of any kind. Statistical errors 
are small, $0.1 \% - 0.3 \%$ in most cases.

As a new application of the method the ground-state
energy and effective polaron mass were calculated 
for the three-dimensional Holstein model [see Figs.~\ref{fig11} 
and \ref{fig2} (triangles)]. At present, three-dimensional 
lattices are out of reach of ED and DMRG methods due to the enormous 
phonon Hilbert space. The present Monte Carlo algorithm treats
all the phonons in an infinite lattice exactly through the
analytical integration so the lattice size is irrelevant.  
Another advantage of the method is its ability to study 
long-range electron-phonon interactions. Indeed, the interaction
enters the formalism only via lattice sums 
$\sum_{\bf m} f_{\bf m}({\bf r}_1) f_{\bf m}({\bf r}_2)$
which can be computed in advance for all ${\bf r}_1 - {\bf r}_2$
and stored for later calculation of the action $A$.
Figures~\ref{fig11} and \ref{fig2} (open squares) show results for
$f_{\bf m}({\bf n}) = \kappa (| {\bf m} - {\bf n} |^2 + 1)^{-3/2}$
in $d=2$ (for this interaction 
$\lambda = 1.742\, \kappa^2/(8tM\omega^2)$).
This form of $f_{\bf m}({\bf n})$ has recently been proposed 
to model the interaction between in-plane holes and 
{\em apical} oxygens in high-$T_c$ superconductors 
\cite{Alexandrov_two}. One can see that at large $\lambda$
the polaron is much lighter than in the Holstein case.
Indeed, a long-range interaction results in ``wider''
polaron paths and, consequently, in a smaller effective
mass by virtue of Eq.~(\ref{three}).
\begin{figure}[t]
\begin{center}
\leavevmode
\hbox{
\epsfxsize=8.4cm
\epsffile{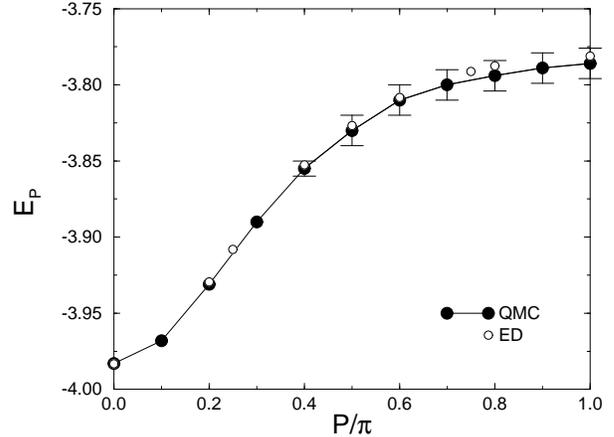}
}
\end{center}
\vspace{-0.5cm}
\caption{ 
The polaron spectrum $E_P$ (in units of $t$) in the one-dimensional
Holstein model for $\bar\omega = 1.0$ and $\lambda=1.75$.
Open circles are exact diagonalization results [20].
}
\label{fig3}
\end{figure}
Finally, I present the polaron {\em spectrum} in the
$d=1$ Holstein model for $\bar\omega=1.0$ and $\lambda=1.75$,
calculated with Eq.~(\ref{two}) (see Fig.~\ref{fig3}).
The spectrum flattens at large momenta as was revealed
in previous numerical studies \cite{Wellein,Stephan_two}. 
The actual energies $E_P$ are in excellent agreement 
with ED results \cite{Wellein_two}.
This demonstrates that a {\em real-time} spectrum can be 
measured directly by an {\em imaginary-time} path-integral QMC.

In conclusion, an efficient continuous-time path-in\-teg\-ral Quantum 
Monte Carlo algorithm for the lattice polaron has been presented. 
It is free from the finite-size and finite-time-step
systematic errors. The method works for infinite lattices and
any range of electron-phonon interaction. It allows for 
exact calculation of the ground-state
energy, effective mass and spectrum of the polaron. 
More technical details will be presented elsewhere.
 
I am very grateful to H.\,Fehske, V.\,Kabanov, W.\,Stephan, and
G.\,Wellein for performing calculations especially to 
compare with results of this work. Useful discussions with
W.\,M.\,C.\,Foulkes are appreciated. This work was supported 
by EPSRC grant No. GR/L40113.

\end{document}